\documentclass[RNAAS]{aastex62}

\usepackage{amsmath}

\graphicspath{{./}{figures/}}

\begin{document}

\title{Galaxy Distribution Incompleteness Testing Using Self-Organizing Maps}

\email{imcmahon@andrew.cmu.edu, markusr@andrew.cmu.edu, rmandelb@andrew.cmu.edu}

\author{Isaac McMahon}
\affil{McWilliams Center for Cosmology, Department of Physics, Carnegie Mellon University, Pittsburgh, PA, 15213}

\author{Markus Michael Rau}
\affil{McWilliams Center for Cosmology, Department of Physics, Carnegie Mellon University, Pittsburgh, PA, 15213}
\affil{High Energy Physics Division, Argonne National Laboratory, Lemont, IL 60439}

\author{Rachel Mandelbaum}
\affil{McWilliams Center for Cosmology, Department of Physics, Carnegie Mellon University, Pittsburgh, PA, 15213}

\section{Abstract}
The calibration of redshift distributions for photometric samples using spectroscopic surveys is plagued by the difficulty in modelling the selection functions of spectroscopic surveys. In this work, we analyse how these selection functions impact redshift inference and quantify the induced biases using local calibration tests in photometry space. The study is carried out using simulations that mimic the radial selection function of a spectroscopic survey and an accompanying mock catalog of a photometric galaxy survey catalog. We use a self-organizing map to partition the photometry space and perform a local $\chi^2$ test to study the probability calibration of redshift inferences that use the spectroscopic data for calibration. The goal of this work is to investigate the effect of uncorrected selection functions in the calibration data on redshift prediction accuracy and critically discuss mitigation methods. In particular we test culling-based bias correction techniques, which aim to remove redshift calibration biases by identifying regions in photometry with few spectroscopic calibration data, and propose avenues for future research. We found that removing regions in color-magnitude space that are underpopulated with spectroscopic calibration data does not remove all biases in redshift inference induced by the selection function. 


\section{Introduction} 
With the advent of precision cosmology driven by imaging surveys like the Dark Energy Survey \citep[DES; e.g.,][]{2018ApJS..239...18A}, the Kilo-Degree Survey  \citep[KiDS; e.g.,][]{2017MNRAS.465.1454H}, the Hyper Suprime-Cam survey \citep[HSC; e.g.,][]{2018PASJ...70S...4A}, the Rubin Observatory Legacy Survey of Space and Time  \citep[LSST; e.g.,][]{2019ApJ...873..111I}, the Roman Space Telescope High Latitude Imaging Survey \citep[HLIS; e.g.][]{2015arXiv150303757S} and Euclid \citep[e.g.][]{2011arXiv1110.3193L}, the accurate estimation and calibration of the galaxies' redshifts using photometric information becomes increasingly important. In particular, while measures of structure growth such as weak gravitational lensing do not require precise estimates of redshift for each galaxy, they do require accurate estimation of the redshift distributions of galaxy samples \citep{2015RPPh...78h6901K,2018ARA&A..56..393M}. 

Traditionally these photometric redshifts have been calibrated by direct comparison with spatially overlapping spectroscopic samples.  
Photometric redshifts are calibrated using either spatial information via cross-correlation methods  \citep[e.g.][]{2008ApJ...684...88N, 2013arXiv1303.4722M, 2013MNRAS.433.2857M, 2016MNRAS.462.1683S, 10.1093/mnras/stx691, Morrison2016, 2017arXiv171002517D, 2018MNRAS.477.1664G, 2020A&A...642A.200V, 2021A&A...647A.124H}, by direct comparison with spatially overlapping spectroscopic samples \citep[e.g.][]{2009A&A...500..981C, 2010A&A...523A..31H, 2013ApJ...775...93D, 2018PASJ...70S...9T, 2016PhRvD..94d2005B} or by comparison/inference using narrow-band photometric data \citep[see e.g.][]{2019MNRAS.489..820B}.  The reliance on spectroscopic calibration data in both prediction and calibration has been called into question \citep[e.g.][]{2016PhRvD..94d2005B, schmidt2020evaluation, 2020MNRAS.496.4769H}; however, spectroscopic observations are still crucial for the calibration of the color-redshift relation for current and upcoming photometric survey programs \citep[see e.g.][]{2022arXiv220601620S}. Spectroscopic surveys have complex selection functions that result in  different magnitude-redshift mappings in comparison with photometric surveys. While it is possible to identify and correct for selection functions in magnitude space, this might not be the case for pure line-of-sight selection functions\footnote{See \citet{2016PhRvD..94d2005B} Appendix D for an extreme case.}. If unaccounted for, both types of selection function can bias ensemble redshift estimation and validation using spectroscopic samples. As shown in \citet{schmidt2020evaluation}, \citet{2021arXiv210210473Z}, and \citet{2022arXiv220514568D}, performing local tests for redshift calibration in magnitude space is essential, as selection function-induced redshift calibration biases can `cancel out' in magnitude space.  

In this work we study this effect using simulations of spectroscopic incompleteness by \citet{2020MNRAS.496.4769H}. We add to \citet{2020MNRAS.496.4769H} a statistical significance test for incompleteness localized in magnitude space, and a global test that combines the local tests to obtain a single $p$-value. We also test the ability to correct for selection effects in redshift by excluding parts of the magnitude space, similar to the approach of  \citet{Masters_2015}. To this end, we partition the magnitude space of the mock catalog of the photometric survey photometry using a self-organizing map (SOM). We then predict photometric redshift distributions of individual galaxies using the simulated spectroscopic calibration dataset. After populating the SOM with these predictions, we apply a local $\chi^2$ test and consider several calibration scenarios to investigate the aforementioned effects.  The research note concludes with a discussion of areas for future work.

\section{Dataset}
\citet{2020MNRAS.496.4769H} simulates a realistic selection function of a spectroscopic survey to investigate the impact of spectroscopic incompleteness on Dark Energy Survey science. The mock catalogs consist of a simulated photometric sample that mimicks DES Y1 survey data \citep{2018MNRAS.478..592H, 2019PhRvD..99l3505A}, based on the Buzzard simulations \citep{2019arXiv190102401D}. The spectroscopic mock catalog mimicks the four spectroscopic datasets VVDS-Deep, VVDS-Wide, VIPERS and zCosmos. The spectroscopic selection functions of these surveys are mimicked by simple color cuts and a spatial selection that matched the real spectroscopic surveys. To mimick a realistic redshifting procedure, \citet{2020MNRAS.496.4769H} determine the redshifts by a combination of a cross correlation technique and manual redshifting. The manual redshifting is performed on a small subsample of galaxies by a set of human experts that determine quality flags for the spectral fits. These flags are then mapped to the full sample using a Machine Learning (ML) algorithm. For further details on the data generation process, we refer readers to \citet{2020MNRAS.496.4769H}. The simulated mock catalog provided to us contains spectroscopic redshifts, subject to incompleteness, as well as the true redshifts from the simulations. Furthermore we have the photometry in the DES filter set for all galaxies.  In the following discussion, we denote the dataset (with/without) the  spectroscopic selection function as (spectroscopic/target) samples. The (spectroscopic/target) samples contain (73318/134155) galaxies, respectively. The faintest magnitudes in the $i$-band in the (spectroscopic/target) samples are (24.8/24.7), with median magnitude (21.8/23.0). The spectroscopic and target samples have nearly identical maximum (1.98/2.00) and median (0.58/0.58) redshifts.

\section{Methodology}
\subsection{Photometric Redshift Estimation}
We use the classification-based conditional density estimate described in \citet{2015MNRAS.452.3710R} to estimate individual galaxy redshift distributions. The individual galaxy redshift distribution is parametrized using a histogram model with $M$ histogram bins.
\begin{equation}
    \hat{p}(z | \mathbf{m}) = \sum_{i=1}^{M} \pi_i(\mathbf{m}) \mathbf{1}(z \in [z_L^i, z_R^i]) \, ,
    \label{eq:cond_distribution_para}
\end{equation}
where $\mathbf{m}$ denotes the photometry (the magnitudes of the galaxy in a filter set), 
$[z_L^i, z_R^i]$ denotes the redshift bin edges of the parametrization, and $\pi_i(\mathbf{m})$ denotes the magnitude-dependent histogram heights. The term $\mathbf{1}(z \in [z_L^i, z_R^i])$ is zero if the redshift $z$ lies in the interval defined by $[z_L^i, z_R^i]$ and zero otherwise. The magnitude dependence of $\pi_i(\mathbf{m})$ is learned using a Random Forest Classifier with cross-entropy loss function using a training set with accurate redshift information. If not stated otherwise, we will train the aforementioned Random Forest Classifier using the spectroscopic mock simulation data and apply the model to the simulated photometric dataset. The classification is done with respect to the redshift bins defined in Eq.~\eqref{eq:cond_distribution_para}. The classifier used is implemented using the {\sc sklearn} package \citep{scikit-learn}. Here we use it in its default configuration with 100 training bins. 
We define $z_{\rm phot}$ of each galaxy to be the mean of the conditional distribution estimate derived for that galaxy.
We obtain a mean (standard deviation) of the residual $z_{\rm phot} - z_{\rm spec}$ of 0.0039 (0.14), which is consistent with the performance of comparative methods in \citet{2014MNRAS.445.1482S} listed in Tables~6-8. We note that we do not use outlier removal in the quoted metrics.  

\subsection{Selection Functions}
\paragraph{Random Subsampling} In ideal random subsampling, the training and test sets are drawn from the same parent population. The joint distribution $p(z, \mathbf{m})$ of redshift $z$ and magnitude $\mathbf{m}$ are therefore equal between the training and test set and we would have 
\begin{equation}
    \begin{split}
    p_{\rm test} (z | \mathbf{m}) &= p_{\rm train} (z | \mathbf{m}) \\
    p_{\rm test}(\mathbf{m}) &= p_{\rm train}(\mathbf{m})  \, .
    \end{split}
\end{equation}
In practise this ideal setup is often not achievable in the context of redshift estimation due to instrumental limitations and targeting strategies of the spectroscopic survey. As a result, these selection functions often lead to biases in parameter inference and interpretation in the context of photometric redshift inference. In this context we identify two relevant types of selection functions: 

\paragraph{Covariate Shift} If the test set differs from the training set in the marginal distribution of its covariates, i.e. photometry\footnote{In this note, the term `photometry' can refer to colors, magnitudes or fluxes. Our inference is based on the galaxy magnitudes, so we use the more general term `photometry' and magnitude interchangeably in this note.},  
but not in the conditional distribution of redshift 
given photometry,  
i.e. 
\begin{equation}
\begin{split}
    p_{\rm test} (z | \mathbf{m}) &= p_{\rm train} (z | \mathbf{m}) \\
    p_{\rm test}(\mathbf{m}) &\neq p_{\rm train}(\mathbf{m})  \, ,
    \end{split}
\end{equation}
we refer to this as a covariate shift scenario. 
For more details we refer to the substantial statistics literature on covariate shift corrections \citep[for a review see, e.g.,][]{10.5555/2209761}. 
Spectroscopic surveys observe the galaxy spectrum with much larger resolution than their photometric counterparts. Thus, observing galaxies to faint magnitudes in a spectroscopic survey requires much longer exposure times as compared with photometric surveys. The spectroscopic sample, therefore, `thins out' at faint magnitudes, which implies a covariate shift between the training sample, that is matched to the spectroscopic sample, and the test sample that is given by the photometric sample. 

\paragraph{Line-of-sight selection function} 
If the conditional distributions of redshift 
given the photometry of the training and test sets differ, i.e. 
\begin{equation}
     p_{\rm test} (z | \mathbf{m}) \neq p_{\rm train} (z | \mathbf{m}) \, ,
\end{equation}
we are considering a line-of-sight selection effect. For example, this effect can be introduced due to instrumental effects, like a wavelength-dependent signal-to-noise ratio in spectrographs, which can deteriorate spectroscopic redshift measurements if the characteristic features of the spectrum move outside of the instrumental sensitivity window. As a result there may not be accurate spectroscopic redshift measurements available for higher redshifts. A line-of-sight selection is often coupled with a covariate shift. \citet{2020MNRAS.496.4769H} forward model both the spectroscopic target function and selection functions induced by redshift failures. The final catalogs show both a covariate shift and a line-of-sight selection with respect to the test sample that mimics a `DES-like' photometric survey.  We refer the interested reader to \citet{2020MNRAS.496.4769H} for more details on these simulations.   


\subsection{Local Test of Selection Effects}
One of the goals of photometric redshift validation is to measure how well the conditional density predictions of redshift given photometry are calibrated for the galaxies in the sample.  Good calibration is to be understood as correctly predicted frequentist coverage probability, i.e. within a given credibility interval obtained from the $p(z | \mathbf{m})$ estimate, we would expect the predicted number of true redshifts under data replication. In the more traditional approach, this photometric redshift calibration is not quantified as a function of magnitude or photometry. As shown in \citet{schmidt2020evaluation} this can be especially problematic, since miscalibrations across photometry can balance out, leading to wrong conclusions about the actual quality of photometric redshift estimates. Tests for calibration accuracy therefore need to be performed as a function of location in magnitude space. There exists work in the statistics literature that propose tests for this scenario; we would like to especially highlight \citet{2021arXiv210210473Z} and \citet{2022arXiv220514568D}. Specifically, \citet{2022arXiv220514568D} proposed methodology to calibrate individual galaxy redshift estimates as a function of color space if a representative spectrophotometric calibration dataset is available. We complement this work by studying the impact of non-representative spectrophotometric calibration data, i.e., data that is subject to a line-of-sight selection function, on sample redshift distribution inference.

In this research note, we take a simple approach and train a self-organizing map (SOM), a popular dimensionality reduction technique in cosmology and in particular photometric redshift estimation \citep[e.g.][]{Masters_2015,2020A&A...637A.100W, 2021MNRAS.505.4249M},  
on the magnitude space of the spectroscopic and photometric samples. The SOM is a data-driven way of dividing up a sample based on its high-dimensional distribution in magnitude space into 2D cells, each with galaxies of similar photometry. 

We train a SOM using the training set data that consists of the simulated four-band DES photometry, where we use the {\sc MiniSOM} python package \citep{vettigliminisom}. The SOM is initialized with random values, then trained on the four DES-Y1 magnitudes to arrange the galaxies by similarity to each other on the map. We estimate the conditional density estimates 
for each galaxy in the test set and then populate the trained SOM using the test set galaxies and their estimated conditional density estimates. Each SOM cell has then a corresponding training and test set sample as well as a set of conditional density estimates for each test set galaxy in the cell. 

On each of these cells we test if the conditional distribution estimates are well calibrated, where we define correct calibration as uniformly distributed $\text{PIT}(z) = \int_{0}^{z} \mathrm{d}z^{'} \hat{p}(z^{'} | \mathbf{m})$ with respect to the true redshifts $z$. To test the quality of the calibration, we perform a $\chi^2$ test between the distribution of PIT values obtained by evaluating all estimated conditional distributions at the correct redshift, and the uniform distribution. This test is performed for all galaxies in each SOM cell.   
We obtain a $p$-value for each SOM cell that we compare with a Bonferroni-corrected significance level of $\alpha=0.05/N_{\rm SOM}$.  
The Bonferroni correction adapts the significance level for the number of simultaneous hypotheses that are tested to counteract the increased chance to reject the null hypothesis by random chance when carrying out  multiple comparisons. Here $N_{\rm SOM}$ is the number of populated cells and therefore the number of tension tests we perform between the calibration data and the estimated individual galaxy photometric redshift distributions $\hat{p}(z | \mathbf{m})$. 

\subsection{Bias Mitigation using SOM-Cell Culling }
Following \citet{Masters_2015}, we now remove a given fraction of SOM cells with mean spectroscopic redshift values in the calibration sample that have a large statistical error $\sigma_{\rm \left\langle z_{\rm spec} \right\rangle} = \frac{\sigma_{\rm spec}}{\sqrt{N_{\rm spec, cell}}}$  
starting with cells that have the largest error. Here, $\sigma_{\rm spec}$ denotes the standard deviation of true redshift values. We hold the SOM partition and the respective data and $\hat{p}(z | \mathbf{m})$ estimates attached to the SOM cells constant. Given sufficient flexibility of the conditional density estimates and a proportionality between the number of spectroscopic populated SOM cells and the severity of the selection effect, one could expect this procedure to remove cells that would yield especially biased predictions. We note that this is a criterion based on how well the spectroscopic calibration data `populates' the SOM. During SOM-cell culling, `goodness-of-fit' criteria constructed using photometric redshift estimates are not used.

\section{Results}
\begin{figure}[h]
    \centering
    \includegraphics[width=\linewidth]{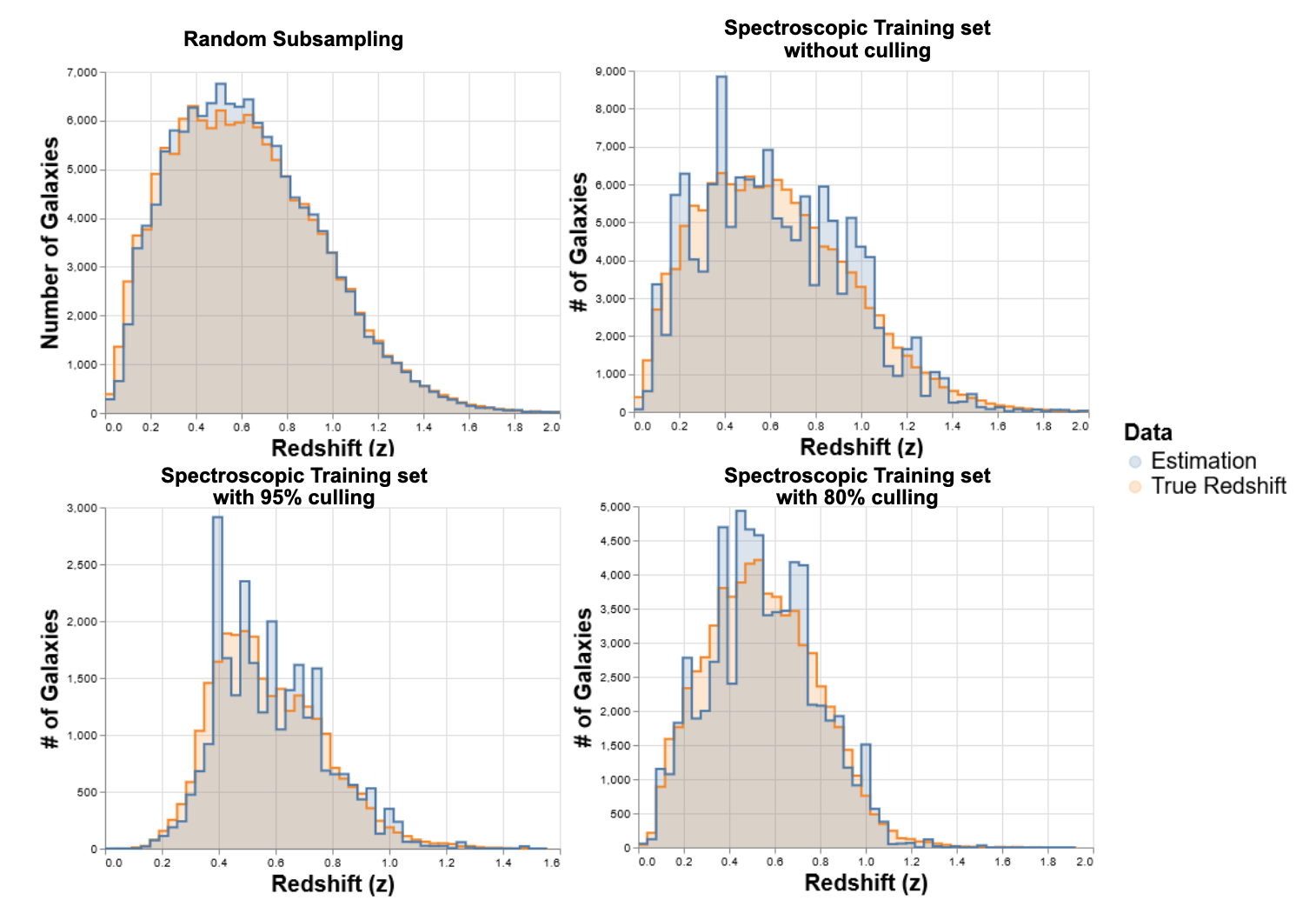}
    \caption{Sample photometric redshift distributions obtained using the true redshifts (orange) and by stacking conditional density estimates (blue). The (top left/top right/bottom left/bottom right) subpanels show the scenarios of a (Random Subsampling/Spectroscopic Training set without culling/Spectroscopic Training set with 95\% culling/Spectroscopic Training set with 80\% culling). 
    The x-axis shows the redshift, the y-axis the number of galaxies in each redshift bin.}
    \label{fig:pop}
\end{figure}

We consider three scenarios that illustrate the effect of spectroscopic incompleteness on photometric redshift predictions: 
\paragraph{Random Subsampling} We construct a disjunct randomly selected subsample by randomly splitting the full test set in half. One part is assigned as the training set, the other as the new test set. We now quantify the performance of the photometric redshift estimates on the test set following the procedures described in the previous sections. Fig.~\ref{fig:pop} top left shows the photometric redshift distributions of the sample of galaxies estimated using the `stacked' conditional photometric redshift estimates with the distribution of true redshift. The stacked distribution is constructed as $\hat{p}(z) = \sum_{i = 1}^{N_{\rm gal}} \hat{p}(z_i | \mathbf{m}_i)$. 
While we note that stacking is not appropriate in general \citep[see e.g.][]{2021PhRvD.103h3502M, 2021arXiv210101184R}, we use it for illustration, so the methodological error can therefore be neglected here. Both distributions visually agree well. To quantify the agreement in magnitude space, we perform the statistical tests described in the previous sections and list the results in Tab.~\ref{tab:samples}. We report the failure rates for the different scenarios in the first column. The number of populated cells after culling is shown in the second column and the failure rates (with Bonferroni corrected $\alpha$) in the third column. The failure rate for `Random Subsampling', as listed in Tab.~\ref{tab:samples}, is  $3.25\%$, which means that our test detects a significant miscalibration in only $3.25\%$ of the SOM cells, which we interpret as model error, i.e. the intrinsic deficiencies in the model training.   
This failure rate serves as the lower bound on the accuracy for the following discussion and will be compared with other results, listed in the first column of Tab.~\ref{tab:samples}, that are additionally impacted by the line-of-sight and covariate shift selection functions. We will consider these scenarios in the following paragraphs.

\paragraph{Spectroscopic Training Set without culling} We use the simulated spectroscopic data as a training set to train a conditional density estimator to predict conditional density estimates for the full test set. Fig.~\ref{fig:pop} top right panel shows the corresponding photometric redshift distributions of the stacked predicted sample redshift distribution and the photometric redshift distribution of the sample of true redshifts. We see that the covariate shift and line-of-sight selection effects in the training data 
deteriorate the agreement between the sample photometric redshift distribution of true redshifts and the predicted photometric sample redshift distribution. Specifically, we see a peaked structure in the latter that is indicative of projection effects, where galaxies are systematically placed in the wrong redshift intervals in undercovered, non-representative areas of magnitude space. The failure rate of SOM cells increases to $21.5\%$ (4th row Tab.~\ref{tab:samples}), as compared with the baseline value for the representative training sample of $3.25\%$ (1st row Tab.~\ref{tab:samples}). This indicates that the selection functions that are present in the training set increase the number of cells that show significant miscalibration.  

\begin{figure}[h]
    \centering
    \includegraphics[width=\linewidth]{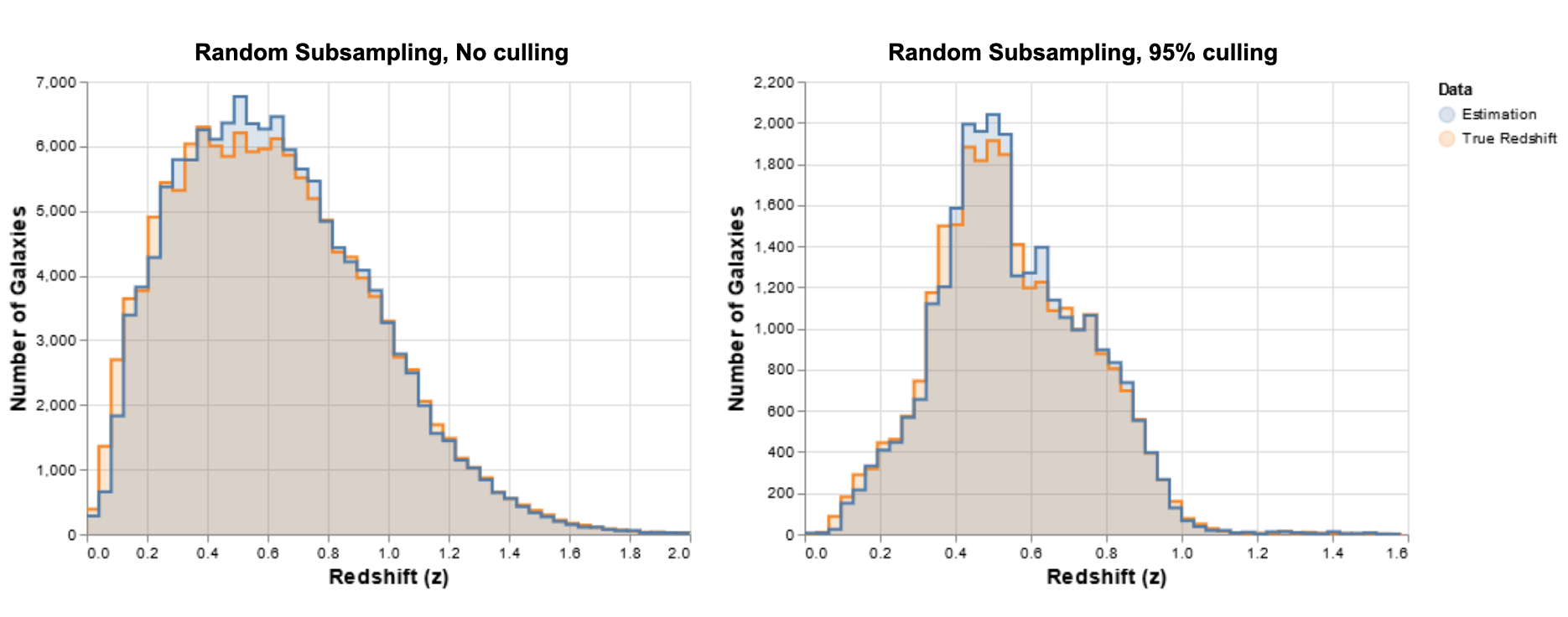}
    \caption{Sample photometric redshift distributions obtained using the true redshifts (orange) and by stacking conditional density estimates (blue). Both panels show the scenario of `Random Subsampling'. The (left/right) panels show the results (without/95\%) culling. The x-axis shows the redshift, the y-axis the number of galaxies in each redshift bin. 
    }
    \label{fig:comparison_representative_training}
\end{figure}

\paragraph{Spectroscopic Training Set with culling} 
Fig.~\ref{fig:pop} (lower left/lower right) shows the sample redshift distributions estimated on the test sample with (95\%/80\%) fraction of culled SOM cells, which corresponds to a fraction of (47\%/19\%) remaining galaxies.    
In both cases, the peaked features persist in the sample redshift distribution estimates, similar to the upper right panel. We can clearly distinguish these effects from the corresponding scenario when we consider a representative training set, also with 95\% culling fraction as shown in Fig.~\ref{fig:comparison_representative_training}. We clearly see that the estimates that use a representative training set, while not optimal, lack large projection peaks in the sample redshift distribution. They can therefore be attributed to the line-of-sight and covariate shift selection functions present in the spectroscopic data that cannot be corrected by culling procedures. This is in agreement with the statistical analysis reported in Tab.~\ref{tab:samples}, where we see an increase in the failure rate when applying culling from $21.5\%$ for the Spectroscopic Training Set with $0\%$ cells culled to ($45\%$/$75\%$) for ($80\%$/$95\%$) cells culled.   

We now study the impact of the culling procedure, that itself is a selection function, on the calibration of the conditional density estimates. Applying SOM-cell culling to the `Random Subsampling' scenario, we report an increase in the failure rates (see Tab.~\ref{tab:samples}) from $3.25\%$ to ($15\%$/$40\%$) for ($80\%$/$95\%$). We interpret this increase in failure rate even for this ideal scenario as follows: The SOM-cell culling procedure imposes a selection function that is not present in the training sample which can lead to mild biases in the trained model. Comparing the `Random Subsampling' scenarios with the `Spectroscopic Training Set' scenarios, we note that the selection functions in the spectroscopic training set further deteriorate the probability calibration. This can be seen by comparing the failure rates between the ($80\%$/$95\%$) 'Random Subsampling' with the ($80\%$/$95\%$) 'Spectroscopic Training Set' scenarios in Tab.~\ref{tab:samples}.  
We can therefore conclude that the culling procedure is not able to correct for the degradation of redshift performance due to the aforementioned line-of-sight and covariate shift selection functions.

\begin{table}[]
    \centering
    \begin{tabular}{|c|c|c|c||c||}
        \hline
        Sample & Cells Culled & Galaxies Remaining & Populated Cells & Failure Rate  \\
        \hline
        \hline
        & 0\% & 100\%  & 400 & 3.25\%  \\
        Random Subsampling & 80\% & 50\% & 80 & 15\%  \\
        & 95\% & 20\% & 20 & 40\%  \\
        \hline
        & 0\% & 100\% & 400 & 21.5\%  \\
        Spectroscopic Training Set & 80\% & 47\% & 80 & 45\% \\
        & 95\% & 19\% & 20 & 75\% \\
        \hline
    \end{tabular}
    \caption{Failure rates of the various analysis scenarios discussed in the test. The (first/second/third/fourth/fifth) columns list the analysis scenario, percentage of SOM cells culled, the percentage of galaxies remaining after culling, the number of populated cells in the SOM, and the rate of galaxies that fail the $\chi^2$ test (Failure Rate) with the Bonferroni correction.}
    \label{tab:samples}
\end{table}

\section{Conclusions and Future Work}
As discussed in this note, spectroscopic calibration data is subject to complex selection functions that limit their usefulness as a reliable source of photometric redshift calibration. We have shown that for broad band photometry, culling techniques in magnitude space are not a reliable way to mitigate these biases. This is likely due to the line-of-sight selection functions that are not necessarily identifiable using photometry alone. We note that this result is clearly survey specific and additional photometric bands might change this assessment. Besides the dependency of our results on the specific set of photometric filters, our methodology to test for incompleteness is not directly applicable to real data, since the true redshift values are unknown. It should be seen as a way to forecast the performance of photometric redshift calibration methodology such as culling using simulated data. This is not a limitation of our testing methodology in particular, as it will similarly apply to all statistical calibration tests for conditional probability density function estimates that utilize a spectrophotometric calibration dataset. In future work, it will be interesting to extend this model testing framework towards photometry space in a posterior predictive testing framework applied to invertible ML models. This can then be applied in conjunction with forward modelling of the galaxy population and photometry and will be interesting for future work.

\begin{acknowledgements}
We thank Will Hartley and Chihway Chang for the use of their simulated galaxy catalog, and Brett Andrews for his feedback.
\end{acknowledgements}

\bibliographystyle{mnras}
\bibliography{bibliography}

\begin{thebibliography}{}
\makeatletter
\relax
\def\mn@urlcharsother{\let\do\@makeother \do\$\do\&\do\#\do\^\do\_\do\%\do\~}
\def\mn@doi{\begingroup\mn@urlcharsother \@ifnextchar [ {\mn@doi@}
  {\mn@doi@[]}}
\def\mn@doi@[#1]#2{\def\@tempa{#1}\ifx\@tempa\@empty \href
  {http://dx.doi.org/#2} {doi:#2}\else \href {http://dx.doi.org/#2} {#1}\fi
  \endgroup}
\def\mn@eprint#1#2{\mn@eprint@#1:#2::\@nil}
\def\mn@eprint@arXiv#1{\href {http://arxiv.org/abs/#1} {{\tt arXiv:#1}}}
\def\mn@eprint@dblp#1{\href {http://dblp.uni-trier.de/rec/bibtex/#1.xml}
  {dblp:#1}}
\def\mn@eprint@#1:#2:#3:#4\@nil{\def\@tempa {#1}\def\@tempb {#2}\def\@tempc
  {#3}\ifx \@tempc \@empty \let \@tempc \@tempb \let \@tempb \@tempa \fi \ifx
  \@tempb \@empty \def\@tempb {arXiv}\fi \@ifundefined
  {mn@eprint@\@tempb}{\@tempb:\@tempc}{\expandafter \expandafter \csname
  mn@eprint@\@tempb\endcsname \expandafter{\@tempc}}}

\bibitem[\protect\citeauthoryear{{Abbott} et~al.,}{{Abbott}
  et~al.}{2018}]{2018ApJS..239...18A}
{Abbott} T.~M.~C.,  et~al., 2018, \mn@doi [\apjs] {10.3847/1538-4365/aae9f0},
  \href {https://ui.adsabs.harvard.edu/\#abs/2018ApJS..239...18A} {239, 18}

\bibitem[\protect\citeauthoryear{{Abbott} et~al.,}{{Abbott}
  et~al.}{2019}]{2019PhRvD..99l3505A}
{Abbott} T.~M.~C.,  et~al., 2019, \mn@doi [\prd] {10.1103/PhysRevD.99.123505},
  \href {https://ui.adsabs.harvard.edu/abs/2019PhRvD..99l3505A} {99, 123505}

\bibitem[\protect\citeauthoryear{{Aihara} et~al.,}{{Aihara}
  et~al.}{2018}]{2018PASJ...70S...4A}
{Aihara} H.,  et~al., 2018, \mn@doi [\pasj] {10.1093/pasj/psx066}, \href
  {https://ui.adsabs.harvard.edu/abs/2018PASJ...70S...4A} {70, S4}

\bibitem[\protect\citeauthoryear{{Bonnett} et~al.,}{{Bonnett}
  et~al.}{2016}]{2016PhRvD..94d2005B}
{Bonnett} C.,  et~al., 2016, \mn@doi [\prd] {10.1103/PhysRevD.94.042005}, \href
  {https://ui.adsabs.harvard.edu/abs/2016PhRvD..94d2005B} {94, 042005}

\bibitem[\protect\citeauthoryear{{Buchs} et~al.,}{{Buchs}
  et~al.}{2019}]{2019MNRAS.489..820B}
{Buchs} R.,  et~al., 2019, \mn@doi [\mnras] {10.1093/mnras/stz2162}, \href
  {https://ui.adsabs.harvard.edu/abs/2019MNRAS.489..820B} {489, 820}

\bibitem[\protect\citeauthoryear{{Coupon} et~al.,}{{Coupon}
  et~al.}{2009}]{2009A&A...500..981C}
{Coupon} J.,  et~al., 2009, \mn@doi [\aap] {10.1051/0004-6361/200811413}, \href
  {https://ui.adsabs.harvard.edu/abs/2009A&A...500..981C} {500, 981}

\bibitem[\protect\citeauthoryear{{Dahlen} et~al.,}{{Dahlen}
  et~al.}{2013}]{2013ApJ...775...93D}
{Dahlen} T.,  et~al., 2013, \mn@doi [\apj] {10.1088/0004-637X/775/2/93}, \href
  {https://ui.adsabs.harvard.edu/abs/2013ApJ...775...93D} {775, 93}

\bibitem[\protect\citeauthoryear{{Davis} et~al.,}{{Davis}
  et~al.}{2017}]{2017arXiv171002517D}
{Davis} C.,  et~al., 2017, arXiv e-prints, \href
  {https://ui.adsabs.harvard.edu/\#abs/2017arXiv171002517D} {p.
  arXiv:1710.02517}

\bibitem[\protect\citeauthoryear{{DeRose} et~al.,}{{DeRose}
  et~al.}{2019}]{2019arXiv190102401D}
{DeRose} J.,  et~al., 2019, arXiv e-prints, \href
  {https://ui.adsabs.harvard.edu/abs/2019arXiv190102401D} {p. arXiv:1901.02401}

\bibitem[\protect\citeauthoryear{{Dey}, {Zhao}, {Newman}, {Andrews}, {Izbicki}
  \& {Lee}}{{Dey} et~al.}{2022}]{2022arXiv220514568D}
{Dey} B.,  {Zhao} D.,  {Newman} J.~A.,  {Andrews} B.~H.,  {Izbicki} R.,   {Lee}
  A.~B.,  2022, arXiv e-prints, \href
  {https://ui.adsabs.harvard.edu/abs/2022arXiv220514568D} {p. arXiv:2205.14568}

\bibitem[\protect\citeauthoryear{{Gatti} et~al.,}{{Gatti}
  et~al.}{2018}]{2018MNRAS.477.1664G}
{Gatti} M.,  et~al., 2018, \mn@doi [\mnras] {10.1093/mnras/sty466}, \href
  {https://ui.adsabs.harvard.edu/\#abs/2018MNRAS.477.1664G} {477, 1664}

\bibitem[\protect\citeauthoryear{{Hartley} et~al.,}{{Hartley}
  et~al.}{2020}]{2020MNRAS.496.4769H}
{Hartley} W.~G.,  et~al., 2020, \mn@doi [\mnras] {10.1093/mnras/staa1812},
  \href {https://ui.adsabs.harvard.edu/abs/2020MNRAS.496.4769H} {496, 4769}

\bibitem[\protect\citeauthoryear{{Hildebrandt} et~al.,}{{Hildebrandt}
  et~al.}{2010}]{2010A&A...523A..31H}
{Hildebrandt} H.,  et~al., 2010, \mn@doi [\aap] {10.1051/0004-6361/201014885},
  \href {https://ui.adsabs.harvard.edu/abs/2010A&A...523A..31H} {523, A31}

\bibitem[\protect\citeauthoryear{{Hildebrandt} et~al.,}{{Hildebrandt}
  et~al.}{2017}]{2017MNRAS.465.1454H}
{Hildebrandt} H.,  et~al., 2017, \mn@doi [\mnras] {10.1093/mnras/stw2805},
  \href {https://ui.adsabs.harvard.edu/abs/2017MNRAS.465.1454H} {465, 1454}

\bibitem[\protect\citeauthoryear{{Hildebrandt} et~al.,}{{Hildebrandt}
  et~al.}{2021}]{2021A&A...647A.124H}
{Hildebrandt} H.,  et~al., 2021, \mn@doi [\aap] {10.1051/0004-6361/202039018},
  \href {https://ui.adsabs.harvard.edu/abs/2021A&A...647A.124H} {647, A124}

\bibitem[\protect\citeauthoryear{{Hoyle} et~al.,}{{Hoyle}
  et~al.}{2018}]{2018MNRAS.478..592H}
{Hoyle} B.,  et~al., 2018, \mn@doi [\mnras] {10.1093/mnras/sty957}, \href
  {https://ui.adsabs.harvard.edu/abs/2018MNRAS.478..592H} {478, 592}

\bibitem[\protect\citeauthoryear{{Ivezi{\'c}} et~al.,}{{Ivezi{\'c}}
  et~al.}{2019}]{2019ApJ...873..111I}
{Ivezi{\'c}} {\v{Z}}.,  et~al., 2019, \mn@doi [\apj]
  {10.3847/1538-4357/ab042c}, \href
  {https://ui.adsabs.harvard.edu/abs/2019ApJ...873..111I} {873, 111}

\bibitem[\protect\citeauthoryear{{Kilbinger}}{{Kilbinger}}{2015}]{2015RPPh...78h6901K}
{Kilbinger} M.,  2015, \mn@doi [Reports on Progress in Physics]
  {10.1088/0034-4885/78/8/086901}, \href
  {https://ui.adsabs.harvard.edu/abs/2015RPPh...78h6901K} {78, 086901}

\bibitem[\protect\citeauthoryear{{Laureijs} et~al.,}{{Laureijs}
  et~al.}{2011}]{2011arXiv1110.3193L}
{Laureijs} R.,  et~al., 2011, arXiv e-prints, \href
  {https://ui.adsabs.harvard.edu/\#abs/2011arXiv1110.3193L} {p.
  arXiv:1110.3193}

\bibitem[\protect\citeauthoryear{{Malz}}{{Malz}}{2021}]{2021PhRvD.103h3502M}
{Malz} A.~I.,  2021, \mn@doi [\prd] {10.1103/PhysRevD.103.083502}, \href
  {https://ui.adsabs.harvard.edu/abs/2021PhRvD.103h3502M} {103, 083502}

\bibitem[\protect\citeauthoryear{{Mandelbaum}}{{Mandelbaum}}{2018}]{2018ARA&A..56..393M}
{Mandelbaum} R.,  2018, \mn@doi [\araa] {10.1146/annurev-astro-081817-051928},
  \href {https://ui.adsabs.harvard.edu/abs/2018ARA&A..56..393M} {56, 393}

\bibitem[\protect\citeauthoryear{Masters et~al.,}{Masters
  et~al.}{2015}]{Masters_2015}
Masters D.,  et~al., 2015, \mn@doi [The Astrophysical Journal]
  {10.1088/0004-637x/813/1/53}, 813, 53

\bibitem[\protect\citeauthoryear{{McQuinn} \& {White}}{{McQuinn} \&
  {White}}{2013}]{2013MNRAS.433.2857M}
{McQuinn} M.,  {White} M.,  2013, \mn@doi [\mnras] {10.1093/mnras/stt914},
  \href {https://ui.adsabs.harvard.edu/\#abs/2013MNRAS.433.2857M} {433, 2857}

\bibitem[\protect\citeauthoryear{{M{\'e}nard}, {Scranton}, {Schmidt},
  {Morrison}, {Jeong}, {Budavari}  \& {Rahman}}{{M{\'e}nard}
  et~al.}{2013}]{2013arXiv1303.4722M}
{M{\'e}nard} B.,  {Scranton} R.,  {Schmidt} S.,  {Morrison} C.,  {Jeong} D.,
  {Budavari} T.,   {Rahman} M.,  2013, arXiv e-prints, \href
  {https://ui.adsabs.harvard.edu/\#abs/2013arXiv1303.4722M} {p.
  arXiv:1303.4722}

\bibitem[\protect\citeauthoryear{{Morrison}, {Hildebrandt}, {Schmidt},
  {Baldry}, {Bilicki}, {Choi}, {Erben}  \& {Schneider}}{{Morrison}
  et~al.}{2017}]{Morrison2016}
{Morrison} C.~B.,  {Hildebrandt} H.,  {Schmidt} S.~J.,  {Baldry} I.~K.,
  {Bilicki} M.,  {Choi} A.,  {Erben} T.,   {Schneider} P.,  2017, \mn@doi
  [\mnras] {10.1093/mnras/stx342}, \href
  {https://ui.adsabs.harvard.edu/abs/2017MNRAS.467.3576M} {467, 3576}

\bibitem[\protect\citeauthoryear{{Myles} et~al.,}{{Myles}
  et~al.}{2021}]{2021MNRAS.505.4249M}
{Myles} J.,  et~al., 2021, \mn@doi [\mnras] {10.1093/mnras/stab1515}, \href
  {https://ui.adsabs.harvard.edu/abs/2021MNRAS.505.4249M} {505, 4249}

\bibitem[\protect\citeauthoryear{{Newman}}{{Newman}}{2008}]{2008ApJ...684...88N}
{Newman} J.~A.,  2008, \mn@doi [\apj] {10.1086/589982}, \href
  {https://ui.adsabs.harvard.edu/\#abs/2008ApJ...684...88N} {684, 88}

\bibitem[\protect\citeauthoryear{Pedregosa et~al.,}{Pedregosa
  et~al.}{2011}]{scikit-learn}
Pedregosa F.,  et~al., 2011, Journal of Machine Learning Research, 12, 2825

\bibitem[\protect\citeauthoryear{Raccanelli, Rahman  \& Kovetz}{Raccanelli
  et~al.}{2017}]{10.1093/mnras/stx691}
Raccanelli A.,  Rahman M.,   Kovetz E.~D.,  2017, \mn@doi [\mnras]
  {10.1093/mnras/stx691}, 468, 3650

\bibitem[\protect\citeauthoryear{{Rau}, {Seitz}, {Brimioulle}, {Frank},
  {Friedrich}, {Gruen}  \& {Hoyle}}{{Rau} et~al.}{2015}]{2015MNRAS.452.3710R}
{Rau} M.~M.,  {Seitz} S.,  {Brimioulle} F.,  {Frank} E.,  {Friedrich} O.,
  {Gruen} D.,   {Hoyle} B.,  2015, \mn@doi [\mnras] {10.1093/mnras/stv1567},
  \href {https://ui.adsabs.harvard.edu/\#abs/2015MNRAS.452.3710R} {452, 3710}

\bibitem[\protect\citeauthoryear{{Rau}, {Morrison}, {Schmidt}, {Wilson},
  {Mandelbaum}  \& {Mao}}{{Rau} et~al.}{2021}]{2021arXiv210101184R}
{Rau} M.~M.,  {Morrison} C.~B.,  {Schmidt} S.~J.,  {Wilson} S.,  {Mandelbaum}
  R.,   {Mao} Y.~Y.,  2021, arXiv e-prints, \href
  {https://ui.adsabs.harvard.edu/abs/2021arXiv210101184R} {p. arXiv:2101.01184}

\bibitem[\protect\citeauthoryear{{Saglia} et~al.,}{{Saglia}
  et~al.}{2022}]{2022arXiv220601620S}
{Saglia} R.,  et~al., 2022, arXiv e-prints, \href
  {https://ui.adsabs.harvard.edu/abs/2022arXiv220601620S} {p. arXiv:2206.01620}

\bibitem[\protect\citeauthoryear{{S{\'a}nchez} et~al.,}{{S{\'a}nchez}
  et~al.}{2014}]{2014MNRAS.445.1482S}
{S{\'a}nchez} C.,  et~al., 2014, \mn@doi [\mnras] {10.1093/mnras/stu1836},
  \href {https://ui.adsabs.harvard.edu/abs/2014MNRAS.445.1482S} {445, 1482}

\bibitem[\protect\citeauthoryear{{Schmidt} et~al.,}{{Schmidt}
  et~al.}{2020}]{schmidt2020evaluation}
{Schmidt} S.~J.,  et~al., 2020, \mn@doi [\mnras] {10.1093/mnras/staa2799},
  \href {https://ui.adsabs.harvard.edu/abs/2020MNRAS.499.1587S} {499, 1587}

\bibitem[\protect\citeauthoryear{{Scottez} et~al.,}{{Scottez}
  et~al.}{2016}]{2016MNRAS.462.1683S}
{Scottez} V.,  et~al., 2016, \mn@doi [\mnras] {10.1093/mnras/stw1500}, \href
  {https://ui.adsabs.harvard.edu/\#abs/2016MNRAS.462.1683S} {462, 1683}

\bibitem[\protect\citeauthoryear{{Spergel} et~al.,}{{Spergel}
  et~al.}{2015}]{2015arXiv150303757S}
{Spergel} D.,  et~al., 2015, arXiv e-prints, \href
  {https://ui.adsabs.harvard.edu/\#abs/2015arXiv150303757S} {p.
  arXiv:1503.03757}

\bibitem[\protect\citeauthoryear{Sugiyama \& Kawanabe}{Sugiyama \&
  Kawanabe}{2012}]{10.5555/2209761}
Sugiyama M.,  Kawanabe M.,  2012, Machine Learning in Non-Stationary
  Environments: Introduction to Covariate Shift Adaptation.
The MIT Press

\bibitem[\protect\citeauthoryear{{Tanaka} et~al.,}{{Tanaka}
  et~al.}{2018}]{2018PASJ...70S...9T}
{Tanaka} M.,  et~al., 2018, \mn@doi [\pasj] {10.1093/pasj/psx077}, \href
  {https://ui.adsabs.harvard.edu/abs/2018PASJ...70S...9T} {70, S9}

\bibitem[\protect\citeauthoryear{Vettigli}{Vettigli}{2018}]{vettigliminisom}
Vettigli G.,  2018, MiniSom: minimalistic and NumPy-based implementation of the
  Self Organizing Map, \url {https://github.com/JustGlowing/minisom/}

\bibitem[\protect\citeauthoryear{{Wright}, {Hildebrandt}, {van den Busch}  \&
  {Heymans}}{{Wright} et~al.}{2020}]{2020A&A...637A.100W}
{Wright} A.~H.,  {Hildebrandt} H.,  {van den Busch} J.~L.,   {Heymans} C.,
  2020, \mn@doi [\aap] {10.1051/0004-6361/201936782}, \href
  {https://ui.adsabs.harvard.edu/abs/2020A&A...637A.100W} {637, A100}

\bibitem[\protect\citeauthoryear{{Zhao}, {Dalmasso}, {Izbicki}  \&
  {Lee}}{{Zhao} et~al.}{2021}]{2021arXiv210210473Z}
{Zhao} D.,  {Dalmasso} N.,  {Izbicki} R.,   {Lee} A.~B.,  2021, arXiv e-prints,
  \href {https://ui.adsabs.harvard.edu/abs/2021arXiv210210473Z} {p.
  arXiv:2102.10473}

\bibitem[\protect\citeauthoryear{{van den Busch} et~al.,}{{van den Busch}
  et~al.}{2020}]{2020A&A...642A.200V}
{van den Busch} J.~L.,  et~al., 2020, \mn@doi [\aap]
  {10.1051/0004-6361/202038835}, \href
  {https://ui.adsabs.harvard.edu/abs/2020A&A...642A.200V} {642, A200}

\makeatother
\end{thebibliography}

\end{document}